\documentclass[12pt]{iopart}

\usepackage{cite}
\usepackage{graphicx}
\usepackage[usenames,dvipsnames]{xcolor}
\usepackage[colorlinks=true, allcolors=blue]{hyperref}
\usepackage{siunitx}

\newcommand{\SiV}{\mathrm{SiV}^-}

\begin{document}

\title[NDs carrying SiV quantum emitters with almost lifetime-limited linewidths]{Nanodiamonds carrying silicon-vacancy quantum emitters with almost lifetime-limited linewidths}

\author[1]{Uwe~Jantzen$^1$, Andrea~B~Filipovski~(Kurz)$^1$\footnote{This author has changed their name since the original version was posted on arXiv.org.}, Daniel~S~Rudnicki$^2$, Clemens~Sch\"afermeier$^3$, Kay~D~Jahnke$^1$, Ulrik~L~Andersen$^3$, Valery~A~Davydov$^4$, Viatcheslav~N~Agafonov$^5$, Alexander~Kubanek$^1$, Lachlan~J~Rogers$^1$, Fedor~Jelezko$^1$}

\address{$^1$ Institute for Quantum Optics and Integrated Quantum Science and Technology (IQ$^\mathrm{st}$), Ulm University, Albert-Einstein-Allee 11, D-89081 Ulm, Germany}
\address{$^2$ Institute of Physics, Jagiellonian University, Lojasiewicza 11, 30-348 Krakow, Poland}
\address{$^3$ Technical University of Denmark, Department of Physics, 2800 Kongens Lyngby, Denmark}
\address{$^4$ L.F. Vereshchagin Institute for High Pressure Physics, Russian Academy of Sciences, Troitsk, Moscow, 142190 Russia}
\address{$^5$ GREMAN, UMR CNRS CEA 6157, Universit\'e F. Rabelais, 37200 Tours, France}
\ead{lachlan.j.rogers@quantum.diamonds}

\begin{abstract}
Colour centres in nanodiamonds are an important resource for applications in quantum sensing, biological imaging, and quantum optics. 
Here we report unprecedented narrow optical transitions for individual colour centres in nanodiamonds smaller than 200\,nm.
This demonstration has been achieved using the negatively-charged silicon vacancy centre, which has recently received considerable attention due to its superb optical properties in bulk diamond.
We have measured an ensemble of silicon-vacancy centres across numerous nanodiamonds to have an inhomogeneous distribution of 1.05\,nm at 5\,K.
Individual spectral lines as narrower than 360\,MHz were measured in photoluminescence excitation, and correcting for apparent spectral diffusion yielded an homogeneous linewidth of about 200\,MHz which is close to the lifetime limit. 
These results indicate the high crystalline quality achieved in these nanodiamond samples, and advance the applicability of nanodiamond-hosted colour centres for quantum optics applications.
\end{abstract}

\submitto{\NJP}

\maketitle

Nanodiamonds (NDs) hosting optically active point defects (``colour centres'') are an important technical material for applications in quantum sensing \cite{ermakova2013detection}, biological imaging \cite{fu2007characterization,tisler2011highly, simpson2014vivo}, and quantum optics \cite{le_floch2014addressing}. 
One colour centre which has attracted recent attention is the negatively charged silicon vacancy ($\SiV$) defect, which consists of a silicon atom taking the place of two adjacent carbon atoms in the lattice \cite{goss1996twelve-line}.
The $\SiV$ centre in diamond has risen to prominence on the basis of its superb spectral properties, including a strong zero-phonon line (ZPL) at \SI{737}{\nano\metre} which contains 70\% of the fluorescence from this colour centre \cite{collins1994annealing}.
In low-strain bulk diamond, the $\SiV$ centre has exhibited lifetime-limited spectral linewidths at 4K with no spectral diffusion \cite{rogers2014multiple}. 
These ideal properties have enabled the efficient production of indistinguishable photons from distinct emitters \cite{sipahigil2014indistinguishable}.
Recent studies in bulk diamond have shown that the electronic spin coherence time in the $\SiV$ centre is fundamentally limited by fast phonon-induced orbital relaxation in the ground state \cite{rogers2014all-optical, jahnke2015electron-phonon}.
Small NDs should impose boundary conditions that prevent the availablilty of phonons at the critical frequency, thereby extending coherence time.
This has increased the motivation to find well-behaved $\SiV$ centres in the nanodiamond environment.

Although $\SiV$ centres have been observed to fluoresce in NDs as small as molecules (\SI{1.6}{\nano\meter}) \cite{vlasov2014molecular-sized}, the ND host has always led to less homogeneous photon emission \cite{neu2011narrowband, neu2011single, grudinkin2012luminescent, neu2013low-temperature}.
Some promising results have been recently reported for larger hybrid nanostructures \cite{zhang2016hybrid}, but the obstacle persists for $\SiV$ applications requiring ND environments.
Here we report unprecedented optical properties of $\SiV$ colour centres hosted in nanodiamonds. 
Individual spectral lines close to the lifetime limit were measured for $\SiV$ centres in  nanodiamonds smaller than \SI{200}{\nano\meter}, representing an improvement of nearly four times over the best $\SiV$ line previously reported for nanodiamonds \cite{muller2014optical}.
Such narrow lines in small nanodiamonds are of interest for a range of applications, including coupling to cavities \cite{riedrich-moller2014deterministic}.

The nanodiamond crystals used in this study were produced using a recently reported synthesis technique \cite{davydov2014production}.
This novel technique directly produces nanometre- and micrometre-sized crystals during high-pressure high-temperature (HPHT) diamond synthesis.
HPHT diamond synthesis reproduces the conditions required for natural diamond formation, where the pressure and temperature make diamond the stable form of carbon.
In contrast to industrial bulk diamond synthesis, the metal catalyst was left out in order to ensure micro- and nanodiamond growth.
Silicon was introduced during the growth process, and incorporated into the diamond crystals to form the negatively-charged silicon vacancy centre.
The narrow spectral lines reported here demonstrate that this direct HPHT synthesis technique is capable of producing nanodiamonds with high crystalline quality, which are therefore a valuable technical material for quantum optics applications.
%
%


\begin{figure}

  \includegraphics[width=\columnwidth]{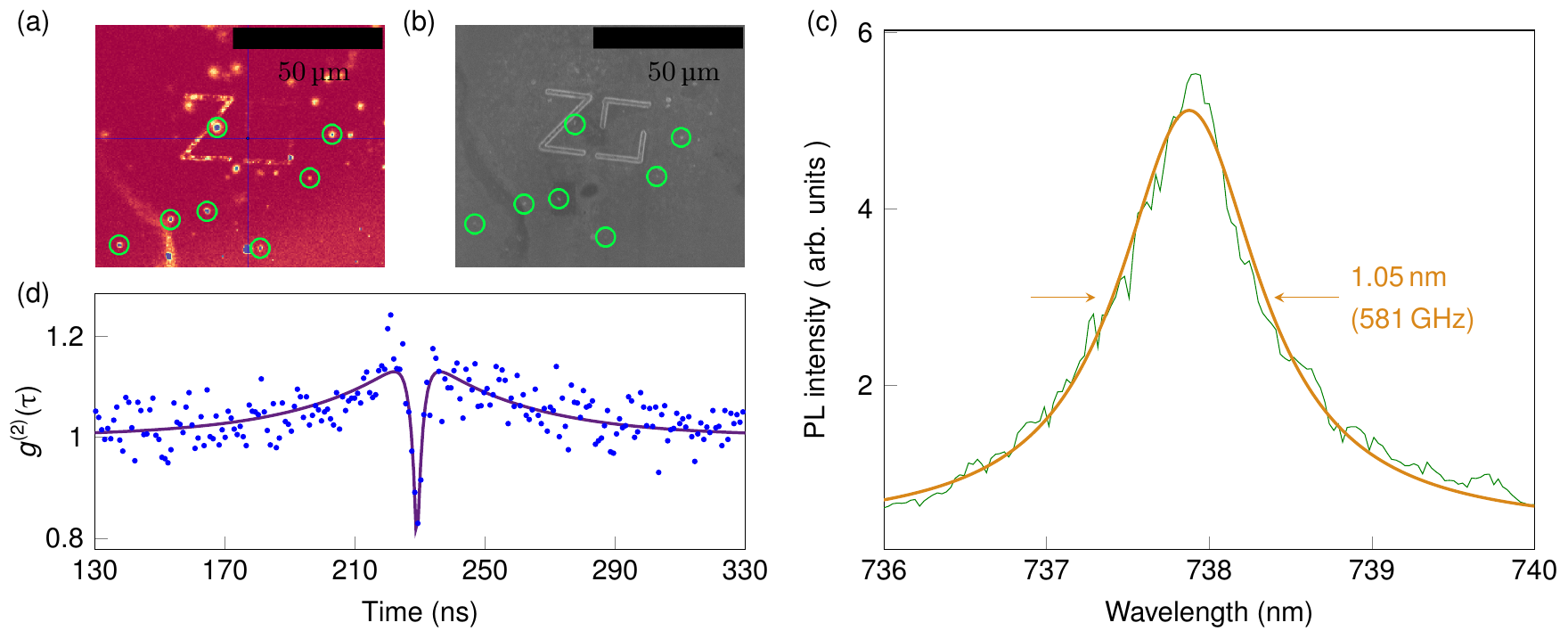}

  \caption{
	  Spectral distribution across multiple nanodiamonds.
	  (a) Fluorescence image of NDs on a diamond substrate near a marker. 
	  Green circles mark spots containing $\SiV$ centres as identified by their fluorescence spectrum.
	  (b) SEM image of same sample region. 
	  The marked spots were identified as NDs (or clusters of NDs) in the SEM and their size determined with an accuracy of $\SI{20}{\nano\meter}$.
	  (c) Photoluminescence spectrum averaged over 7 fluorescent spots containing a total of more than 50 $\SiV$ centres.
	  The illustrated lorentzian fit was used to measure the linewidth.
	  (d) The most visible dip in the $g^{(2)}$ function at a fluorescence spot was only to a depth of \SI{82}{\percent}, corresponding to about 6 emitters (assuming equal brightness).
	  }
  \label{fig:PL}
\end{figure}

The nanodiamonds were suspended in a solution of ultrapure water and ethanol, and were ultrasonicated to disperse the crystals.
This solution was spin coated on a thermally conducting  substrate (type IIa diamond) containing markers to facilitate accurate comparison between confocal fluorescence imaging and scanning electron microscope (SEM) imaging, as shown in \autoref{fig:PL}(a) and (b).
This enabled correlation of the optical spectroscopy with the ND shape and size.
Optical excitation was provided by a continuous-wave \SI{532}{\nano\meter} frequency-doubled diode-pumped solid-state laser.
Fluorescence images and spectra were recorded with a home-built confocal microscope, using an air objective  with NA=0.95.
To resolve the fine-structure of the $\SiV$ the sample was mounted in a helium-flow-cryostat.
The cryostat cold-finger reached a temperature of \SI{5}{\kelvin}, and the thermal conductivity of the substrate suggested that the NDs were at a temperature below \SI{8}{\kelvin}.
Photoluminescence spectra were measured on a spectrometer (grating with 1200 lines/mm) for 7 fluorescent spots containing several $\SiV$ centres. 
The summed ensemble zero-phonon line is shown in \autoref{fig:PL}(c), and was found to have a linewidth of 1.05\,nm (581\,GHz) representing the inhomogeneous distribution across multiple $\SiV$ centres.
This is broader than $\SiV$ ensembles in low-strain bulk diamond which exhibited linewidths of 8\,GHz \cite{dietrich2014isotopically}, but narrower than previously reported ND observations of about 5\,nm (3\,THz) \cite{bohm2015low-temperature}.
It is concluded that the novel HPHT fabrication technique used here is capable of producing NDs with a more uniform crystal lattice than previous fabrication methods.

Due to the diffraction limited resolution of optical microscopy, bright spots in the fluorescence image did not necessarily correspond to individual $\SiV$ centres or even to individual nanodiamonds.
Photon autocorrelation statistics (the $g^{(2)}$ function) are typically used to demonstrate single-emitter detection (where $g^{(2)}(0)<0.5$).
The deepest dip observed here was only to a relative height of $g^{(2)}(0)=\mathrm{0.82}$ as shown in \autoref{fig:PL}(d).
This corresponds to six emitters if they were equally bright, and more than six if some were lying outside the optimum collection region of the confocal microscope.
Most of the fluorescent spots did not produce a measurable dip, suggesting the presence of many $\SiV$ centres.

\begin{figure}
	\centering{\includegraphics[width=0.7\textwidth]{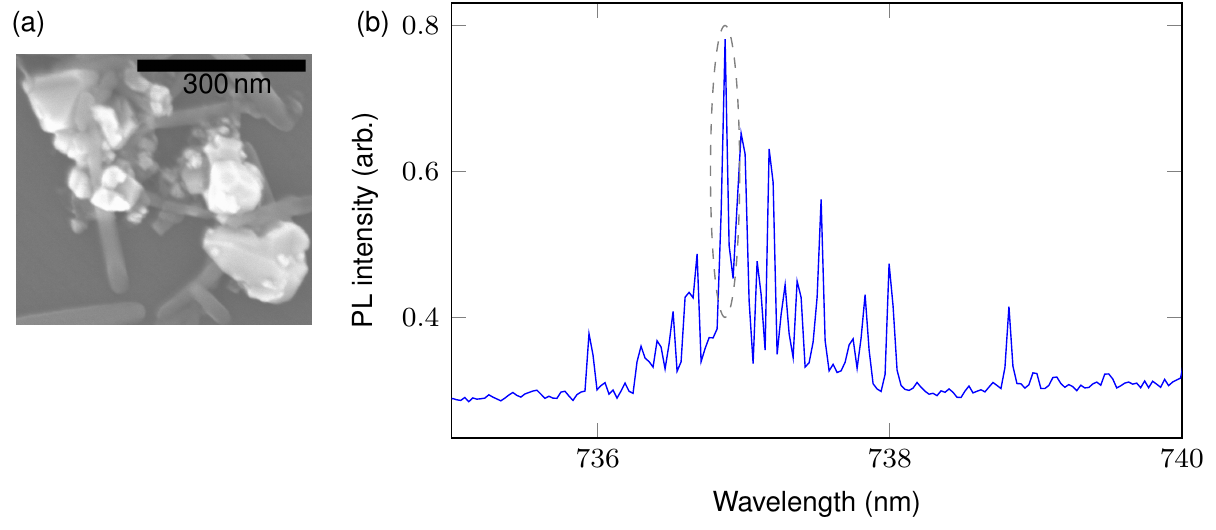}}
	\caption{
	  PL for a single spot in the confocal image.
	  (a) SEM image showing this spot is made up of a number of NDs each smaller than 200\,nm. 
	  It was not possible to associate spectral features with specific NDs within the cluster.
	  (b) PL spectrum recorded for this ND cluster, showing many lines corresponding to numerous $\SiV$ centres.
	  The circled line is the location of the feature examined in PLE in \autoref{fig:single_site_ple}.
	}
  \label{fig:single_spot_pl}
\end{figure}

Despite the ultrasonication used in sample preparation, SEM imaging revealed clustering of NDs as shown in \autoref{fig:single_spot_pl}(a), resulting in more than one ND in the confocal detection spot.
In this case it was not possible to determine which of the clustered NDs contained $\SiV$ centres.
Subsequent to the measurements reported here, some nanodiamonds were spin-coated from a solution of chloroform (with residual ethanol and ultrapure water) following comments in \cite{park2006cavity}, and this reduced clustering but did not eliminate it.
Future experiments may be able to further reduce the clustering of NDs through more advanced preparation techniques.
The PL spectra that exhibited a $\SiV$ ZPL were typically found to contain more than the four-line structure that is expected for a single centre \cite{clark1995silicon, goss1996twelve-line, rogers2014electronic}, as shown in \autoref{fig:single_spot_pl}(b).
This is another indication of the presence of multiple $\SiV$ centres in the fluorescence detection volume.
From the $g^{(2)}$ data and clustering observations we conclude that the inhomogeneous linewidth in \autoref{fig:PL}(c) is from an ensemble of more than 50 $\SiV$ centres.
An interesting implication of the high number-density of $\SiV$ in these small NDs is the reasonable probability of two centres being in close proximity. 
For two nearly resonant centres at close separation, direct dipole-dipole interaction would cause a shifting of the spectral lines.
It is possible that this effect contributes to a broadening of the ensemble linewidth.

\begin{figure}
  \includegraphics[width=\textwidth]{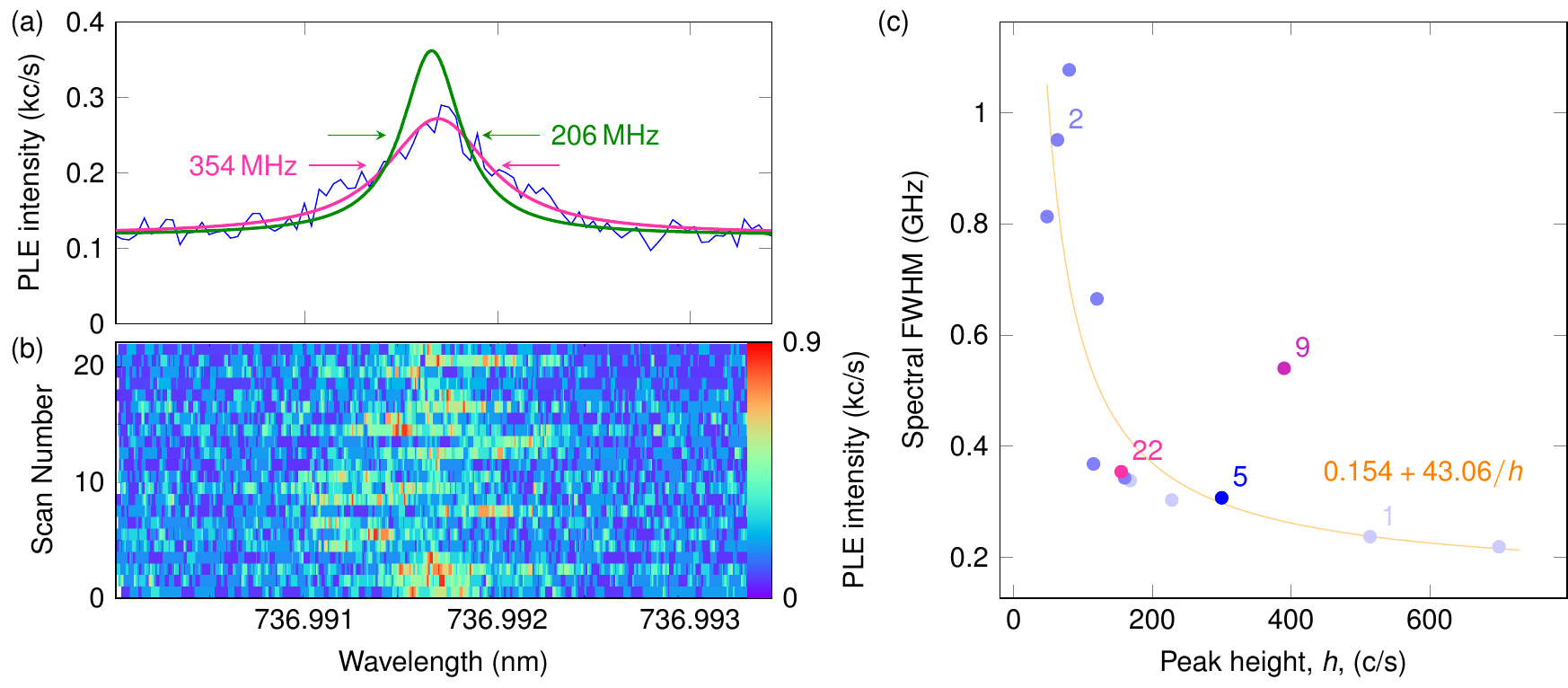}
	\caption{
	  Resonant excitation to probe individual $\SiV$-centres.
	  (a) Photoluminescence excitation (PLE) spectrum of a single transition with averaged linewidth of \SI{354}{\mega\hertz} 
	  (Data in blue, fit to average in purple, spectral diffusion interpretation in green). 
	  (b) The raw data consisted of 22 separate scans, and the line position was found to vary between scans.
	  Interpreting this as spectral diffusion and shifting each scan for correction gave a homogeneous linewidth of \SI{206}{\mega\hertz} (green curve in (b)), which is close to the lifetime limit. 
	  (c) Summary of PLE measurements for 13 spectral lines, showing many with linewidths below 400\,MHz.
	  Pale blue data points are from single-pass scans of the line; middle-blue are from double-pass scans; blue, purple, magenta indicate 5, 9, 22 passes respectively.
	  The magenta point corresponds to the data shown in (a) and (b).
	  The distribution of points closely follows a reciprocal relationship (orange fit excludes the purple spot with 9 passes), indicating that the area under the PLE line was similar for all of these spectra.  
	  It is interpreted that some spots have spectral diffusion over about 1\,GHz at a rate much faster than the measurement (broadening and lowering the PLE peak).
	  }
  \label{fig:single_site_ple}
\end{figure}

Since spatial resolution was unable to isolate individual $\SiV$ centres, resonant excitation techniques were used to allow spectral isolation.
To perform photoluminescence excitation (PLE) spectroscopy a resonant laser was scanned through the zero-phonon line while fluorescence was detected off-resonantly in the 750--810\,nm band (the phonon sideband).
Individual isolated optical transitions were excited in this manner and the spectral linewidths were measured to high precision (the instrument limit of the laser was $<100$\,kHz).
Resonant excitation can lead to power broadening, although this effect was found to be negligible for excitation laser powers below $\SI{4}{\nano\watt}$ entering the microscope objective.
\autoref{fig:single_site_ple}(a) shows a PLE spectrum exhibiting a $\SiV$ linewidth of $\SI{354}{\mega\hertz}$, for a ND below 200\,nm in size from the cluster shown in \autoref{fig:single_spot_pl}(a).
This is considerably narrower than the previous best $\SiV$ lines in NDs of 1.4\,GHz \cite{muller2014optical}.

This excitation line was measured by making multiple scans at 400\,MHz/s and averaging, and it is apparent in \autoref{fig:single_site_ple}(b) that additional information is contained in the individual scans.
The line position was observed to change with time in a manner similar to the spectral diffusion that has been observed for other colour centres in diamond \cite{fu2010conversion}.
Interpreting this behaviour as spectral diffusion and displacing each scan to overlap the peak positions yielded an homogeneous linewidth of $\SI{206}{\mega\hertz}$ as illustrated in \autoref{fig:single_site_ple}(b).
The slope of the $g^{(2)}$ dip in \autoref{fig:PL}(d) indicates that the $\SiV$ centres in these NDs had an excited state decay lifetime of about 1.7\,ns, which is consistent with measurements in bulk diamond \cite{rogers2014multiple}.
This fluorescence lifetime imposes a fourier-transform-limited linewidth of 100\,MHz.
However, it was not possible to reliably identify which of the four ZPL transitions this PLE line was associated with since the distribution of ZPL positions across the ensemble of $\SiV$ centres was far greater than the fine-structure splitting.
At low temperature two of these four transitions are known to be broadened due to thermalisation in the excited state \cite{rogers2014multiple}.
It is therefore difficult in this ND situation to compare the measured linewidth to the lifetime limit in detail, however it is clear that the values are close.

\autoref{fig:single_site_ple}(c) shows a summary of similar PLE measurements made for 13 $\SiV$ lines, illustrating the measured linewidth and peak height.
For technical reasons including spatial drift of the confocal microscope and blinking of the $\SiV$ sites (discussed below), most of the PLE spectra were recorded in far fewer than the 22 passes contained in \autoref{fig:single_site_ple}(b).
It is apparent that single-pass spectra typically gave a narrower linewidth than averages over multiple passes, as expected in the presence of slow spectral diffusion.
However, it is striking that the peaks with the lowest amplitude also had the broadest linewidth.
This is the inverse of the trend expected for the situation of power broadening.
Indeed, the reciprocal relationship illustrated in \autoref{fig:single_site_ple}(c) shows that these PLE lines enclosed essentially the same area and therefore the linewidth variation was not related to excitation intensity. 
It is interpreted that these spectra corresponded to $\SiV$ sites exhibiting spectral diffusion at a rate much faster than the 50\,Hz photon-counting-bins used for these measurements.
Such a situation would mean that even single- or double-pass scans would trace out the averaged ``envelope'' of the rapidly-shifting spectral feature, leading to broader but lower peaks.
Despite this broadening of some of the lines, a clear majority had similar characteristics to the line studied in \autoref{fig:single_site_ple}(a) and (b).

\begin{figure}
  \includegraphics[width=\textwidth]{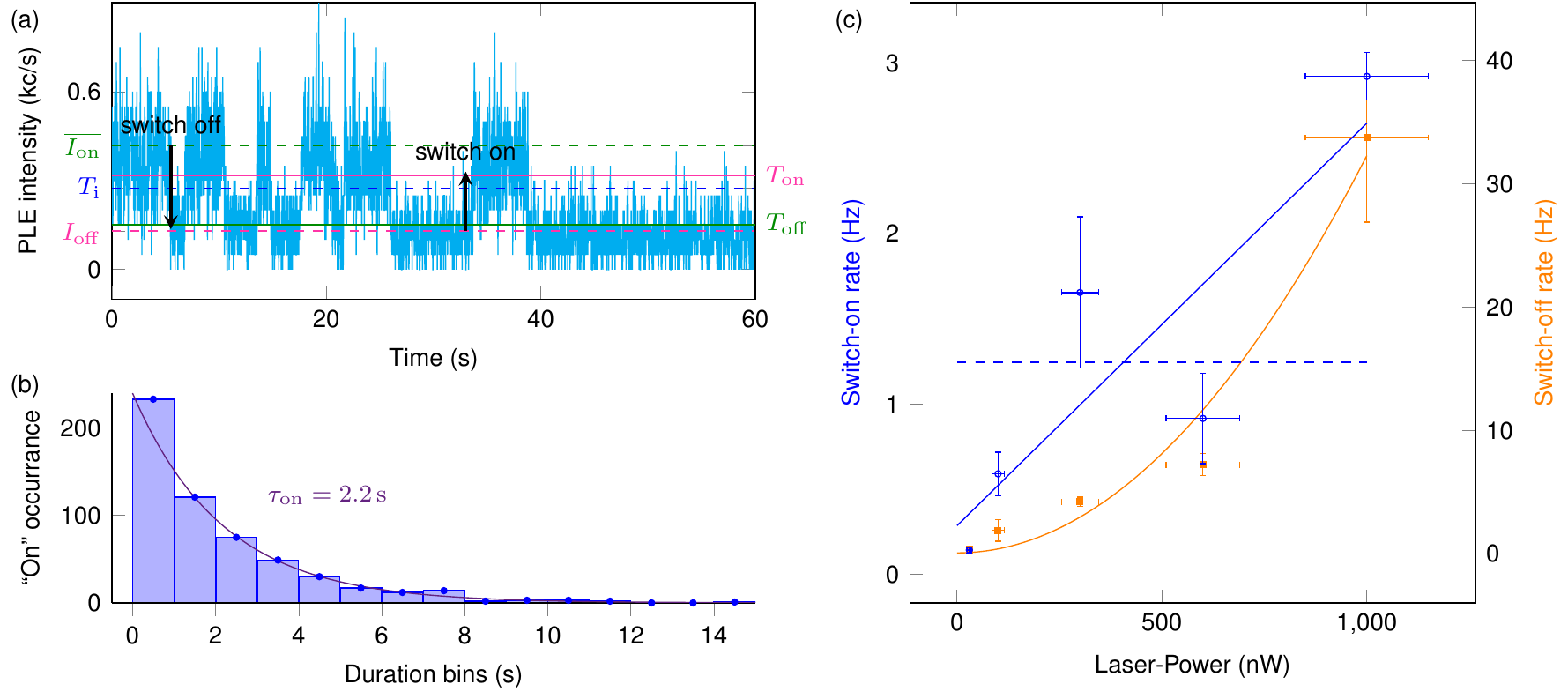}
  \caption{
	  Blinking of fluorescence.
	  (a) Under resonant excitation the $\SiV$ centres were observed to change between two discrete fluorescence levels.  
	  Here the first minute of data for 30\,nW excitation power on a $\SiV$ line at 736.612\,nm is shown as a typical example, with the switching thresholds $T_\mathrm{on}$ and $T_\mathrm{off}$ illustrated for $k=2.5$.
	  The raw measurement data are included in the supplementary data (see blinking\_time\_trace csv files, available from URL).
	  (b) Histograms of duration for the ``on'' and ``off'' intervals gave exponential distributions, from which a characteristic time can be determined (and hence switching rate).
	  The histogram for the ``on'' events is shown for the 30\,nW data and thresholds of $k=2.5$.
	  (c) The ``on'' and ``off'' rates ($R_\mathrm{on}$ and $R_\mathrm{off}$) both increase with excitation laser intensity, with linear and quadratic fits illustrated respectively.
	  The dashed line is the best fit for a constant $R_\mathrm{on}$.
	  The horizontal error bars correspond to a 15\% uncertainty in determining the laser power applied to the $\SiV$ centre, and the vertical error bars represent the uncertainty in judging a switching event.
	  }
  \label{fig:blinking}
\end{figure}

The narrow optical transitions indicate the high crystalline quality of these NDs.
These results are promising for $\SiV$ applications requiring small pieces of diamond.
Unfortunately, a blinking phenomenon was observed in which the fluorescence switched between two discrete levels as shown in \autoref{fig:blinking}(a).
This behaviour is consistent with previous reports of $\SiV$ centres in NDs \cite{neu2012photophysics}, and it introduces challenges in the development of applications involving colour centres in NDs.
In order to obtain more information about the processes responsible for blinking, time series of the fluorescence rate were recorded for a few minutes at various incident laser powers in the range of 30--1000\,nW.
Raw measurement data are included in the supplementary data (see blinking\_time\_trace csv files, available from URL).
The laser frequency was chosen to maximise the fluorescence, indicating resonance with the $\SiV$ optical transition.
The switching rates ($R_\mathrm{on}$, $R_\mathrm{off}$) were determined from the typical durations of ``on'' and ``off'' events in these time series.

A double-threshold technique was used to identify the ``on'' and ``off'' events, since in general the switching contrast was smaller than the noise amplitude as seen in \autoref{fig:blinking}(a).
In this technique the switch-on threshold was higher than the switch-off threshold.
A preliminary threshold $T_\mathrm{i}$ was manually chosen to start this process, shown by the dashed blue line in \autoref{fig:blinking}(a).
The mean $\overline{I_\mathrm{off}}$ and standard deviation $\sigma_\mathrm{off}$ were calculated for all data points below $T_\mathrm{i}$, and $\overline{I_\mathrm{on}}$ and $\sigma_\mathrm{on}$ were determined for the data points above $T_\mathrm{i}$.
Actual switching thresholds were then calculated as 
$T_\mathrm{on}=\overline{I_\mathrm{off}} + k \sigma_\mathrm{off}$
and
$T_\mathrm{off}=\overline{I_\mathrm{on}} - k \sigma_\mathrm{on}$
for some constant $k$.
These thresholds mean that a switch of state is identified only if the signal deviates from the current state by more than $k$ times the standard deviation (noise level) in the current state.

The fundamental uncertainty in extracting ``on'' and ``off'' durations from the blinking time traces arose because short switching events may be indistinguishable from noise spikes.
This is directly related to the strictness of the switching thresholds $T_\mathrm{on}$ and $T_\mathrm{off}$, which are determined by the constant $k$.
Various thresholds beginning at $k=2$ were used, with $k$ increasing to the point where the threshold lost meaning (when $T_\mathrm{off} < \overline{I_\mathrm{off}}$, meaning that the switch-off threshold level went below the mean ``off'' count-rate).
Because $k$ was varied over a broad range (typically up to about $k=3$) it was not important that the means and standard deviations arose from $T_i$ rather than the actual $T_\mathrm{on}$ and $T_\mathrm{off}$.
For each $k$-value histograms were produced for the duration of on and off intervals, and characteristic time-constants were obtained from fitting the histograms with exponential decay functions as illustrated in \autoref{fig:blinking}(b).  
The switching rates ($R_\mathrm{on}$, $R_\mathrm{off}$) were taken as the reciprocal of the characteristic duration of off and on intervals respectively, and are shown as a function of excitation intensity in \autoref{fig:blinking}(c).
The data points and vertical error bars in \autoref{fig:blinking}(c) represent the mean and standard deviation of the results from the range of plausible thresholds (the range of $k$-values).
Although the uncertainty margins are high, it is clear that both rates vary with applied laser power.

Precise identification of these switching processes is challenging and left open for further investigations, however our observations exclude a few obvious candidates. 
The blinking was observed using resonant excitation, which is capable of optically pumping the system to a ``dark'' ground state from which all excitation transitions are non-resonant. 
$\SiV$ has an orbital degeneracy in its ground state, providing a potential dark state, but if this were the cause of the blinking then the switch-on rate $R_\mathrm{on}$ would correspond to the orbital relaxation process in the ground state and should therefore be independent of excitation intensity.
The best fit for such a case is illustrated in \autoref{fig:blinking}(b) by a dashed line, and it is excluded by the data.
In analogy with the nitrogen vacancy $\mathrm{NV}^-$/$\mathrm{NV}^0$ system, photo-ionisation would provide another potential ``dark'' state leading to blinking.
The neutral $\mathrm{SiV}^0$ centre is known in diamond, and has been attributed to a zero-phonon line at 946\,nm.
The 736\,nm resonant excitation for $\SiV$ would therefore be far from resonant to $\mathrm{SiV}^0$, and is unlikely at the low intensities used here to be capable of exciting appreciable photo-ionisation back to the negative charge state.
Unlike the nitrogen vacancy centre, the neutral charge state $\mathrm{SiV}^0$ is too weakly fluorescent to be detectable at the single-site level, making it impossible to check for its presence in the fluorescence spectra measured here.

It has been argued that the $\SiV$ centre is remarkably insensitive to strain and electric field perturbations \cite{rogers2014multiple, sipahigil2014indistinguishable}, but the shielding effects of symmetry should be reduced as the centre becomes more distorted.
In nanodiamonds external charge fluctuations may well be ``visible'' to the $\SiV$ centres, and these are a plausible cause for the spectral diffusion and blinking reported here.  
Any surface chemistry which is photo active would account for the increasing blinking rates with higher excitation intensities.
It has been shown that surface treatment can control blinking of fluorescent colour centres \cite{bradac2010observation} and it is expected that future work in this direction may improve the performance of $\SiV$ centres in NDs.
In fact, these surface effects could be the origin of both the spectral diffusion and the blinking.  
It was not possible to directly compare the spectral diffusion process with the observed blinking because power broadening masks the diffusion effect.
%



In conclusion, we have measured the narrowest $\SiV$ spectral lines in NDs of \SI{354}{\mega\hertz}, and this can be reduced to \SI{200}{\mega\hertz} after correcting for spectral diffusion. 
This is close to the transform limit, and suggests that these direct-HPHT synthesised NDs have a crystal quality that surpasses the NDs used in previous $\SiV$ experiments.
This material is therefore uniquely attractive for use in quantum optics applications, including cavities.
Existing limitations due to blinking effects and spectral diffusion are likely due to interaction with other defects at the surface of the NDs, and this problem is fundamental to all colour centres close to the diamond surface (regardless of crystalline quality).
This should be tackled by surface treatment.
%


\section*{Acknowledgements}
The authors thank W Gawlik for helpful discussions. 
Experiments performed for this work were operated using the Qudi software suite, available from https://github.com/Ulm-IQO/qudi .
This work was funded from ERC project BioQ, EU projects SIQS and EQUAM,  DFG (SFB/TR21 and FOR 1493), Volkswagenstiftung and BMBF.
ABK and AK acknowledge support of the Carl-Zeiss Foundation.
AK also acknowledges support from Wissenschaftler-R\"uckkehrprogramm GSO/CZS, IQST, and DFG. 
DR acknowledges support of the NATO 'Science for Peace' grant (CBP.MD.SFP 983932) and KNOW project. 
VD thanks the Russian Foundation for Basic Research (Grant No. 15-03-04490) for financial support.

\section*{Author Contributions}
UJ, AF, DR, CS, KJ, and LR performed the measurements, which were conceived by LR, AK, and FJ. 
VD and VA supplied the nanodiamond samples.
The manuscript was written by UJ, AF, and LR, and all authors discussed the results and commented on the manuscript.

\section*{References}
\bibliography{siv_nd} 
\bibliographystyle{unsrt} 

\end{document}